\pdfoutput=1
\documentclass[10pt,twocolumn]{article}

\usepackage{arxiv}
\usepackage{amsmath,amssymb}
\usepackage{booktabs}
\usepackage{tabularx}
\usepackage{array}
\usepackage{graphicx}
\usepackage{microtype}
\usepackage[numbers,sort&compress]{natbib}
\usepackage{enumitem}
\usepackage{xcolor}
\usepackage{caption}
\usepackage{hyperref}
\usepackage{balance}
\graphicspath{{figures/}}
\usepackage{cuted}

\hypersetup{
  hidelinks,
  pdftitle={Geography as the Organizing Grammar of Geospatial Models},
  pdfauthor={Rajiv Ranjan and Shashank Tamaskar}
}

\newcommand{\proposition}[1]{%
  \subsubsection*{Proposition #1}%
  \addcontentsline{toc}{subsubsection}{Proposition #1}}
\newcommand{\reviewquestion}[1]{%
  \subsubsection*{Review Question #1}%
  \addcontentsline{toc}{subsubsection}{Review Question #1}}
\newcommand{\answer}{\par\noindent\textbf{Answer.}\ }

\title{\textbf{Geography,as the Grammar of Geospatial Models}}

\author{
  Rajiv Ranjan \qquad Shashank Tamaskar\\
  \normalsize Department of Robotics and Cyber Physical Systems\\
  \normalsize Plaksha University, Mohali, India\\
  \normalsize
  \texttt{rajiv.ranjan@plaksha.edu.in} \qquad
  \texttt{shashank.tamaskar@plaksha.edu.in}
}

\date{}

\begin{document}
\maketitle

\begin{strip}
\centering
\includegraphics[
    width=0.98\textwidth,
    keepaspectratio
]{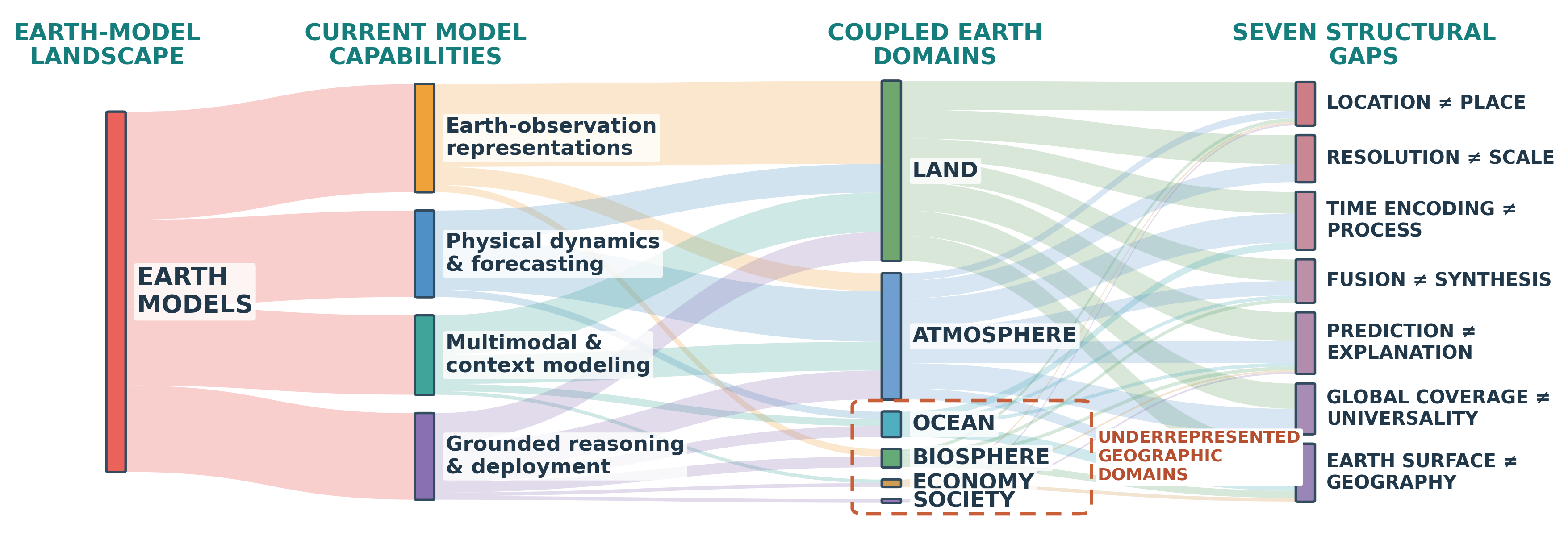}
\end{strip}

\begin{figure*}[!h]
    \centering
    \includegraphics[
        width=0.98\textwidth,
        keepaspectratio
    ]{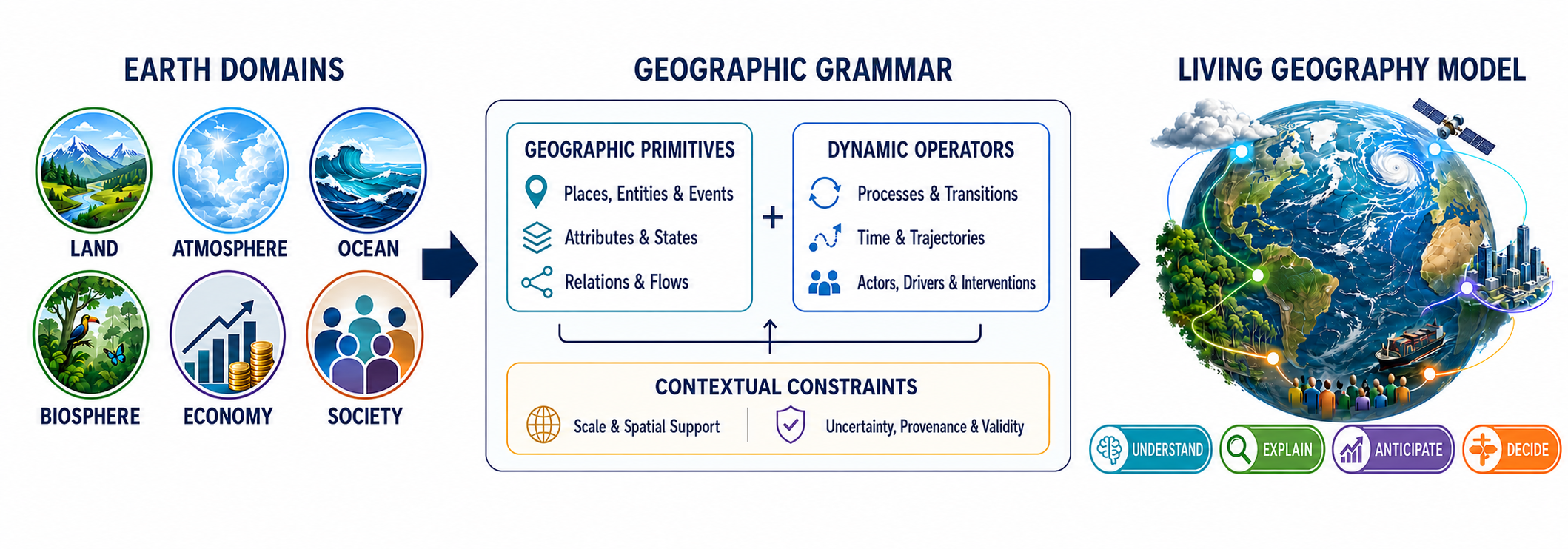}
    \caption{\textbf{Geography as the organizing grammar of intelligent Earth
    models.} Observations from land, atmosphere, ocean, biosphere, economy,
    and society are organized through geographic primitives, dynamic
    operators, and contextual constraints. Their integration enables a
    Living Geography Model capable of understanding, explaining,
    anticipating, and supporting decisions about coupled Earth systems.}
    
    \label{fig:geographic-grammar-framework}
\end{figure*}

\begin{abstract}
GeoAI is transforming Earth observations into reusable
embeddings, physical forecasts, multimodal representations, and automated
reasoning systems. Yet georeferenced data and global coverage do not themselves
constitute geographic intelligence. Contemporary Earth models often conflate
location with place, resolution with scale, temporal encoding with process,
fusion with synthesis, and prediction with explanation. They also privilege
terrestrial surfaces and atmospheric fields, while ocean dynamics,
biogeographic organization, economic networks, institutions, and human geography
remain fragmented across specialist systems. These systemic gaps help explain
the persistent divide between technical progress and transferable,
decision-relevant, real-world implication. This critical integrative review positions geography as the organizing grammar
of intelligent Earth/geospatial models. This grammar can be formalized through three components: geographic primitives representing places and their relationships; dynamic operators capturing processes, trajectories, drivers, and interventions; and contextual constraints governing scale, uncertainty, provenance, and validity.
 Seven propositions distinguish
geographic intelligence from geospatial pattern recognition and translate this
framework into testable requirements. We conclude by proposing Living Geography Models: federated, continually
updated, multiscale systems that connect land, atmosphere, ocean, biosphere,
economy, and society. Such models should move beyond describing where
phenomena occur to explaining how places are connected, why they change, what
may happen next, and how interventions may alter their futures.
\end{abstract}

\keywords{Geography; GeoAI; Earth foundation models; geospatial reasoning}

\begin{figure*}[t]
  \centering
  \includegraphics[width=\textwidth]{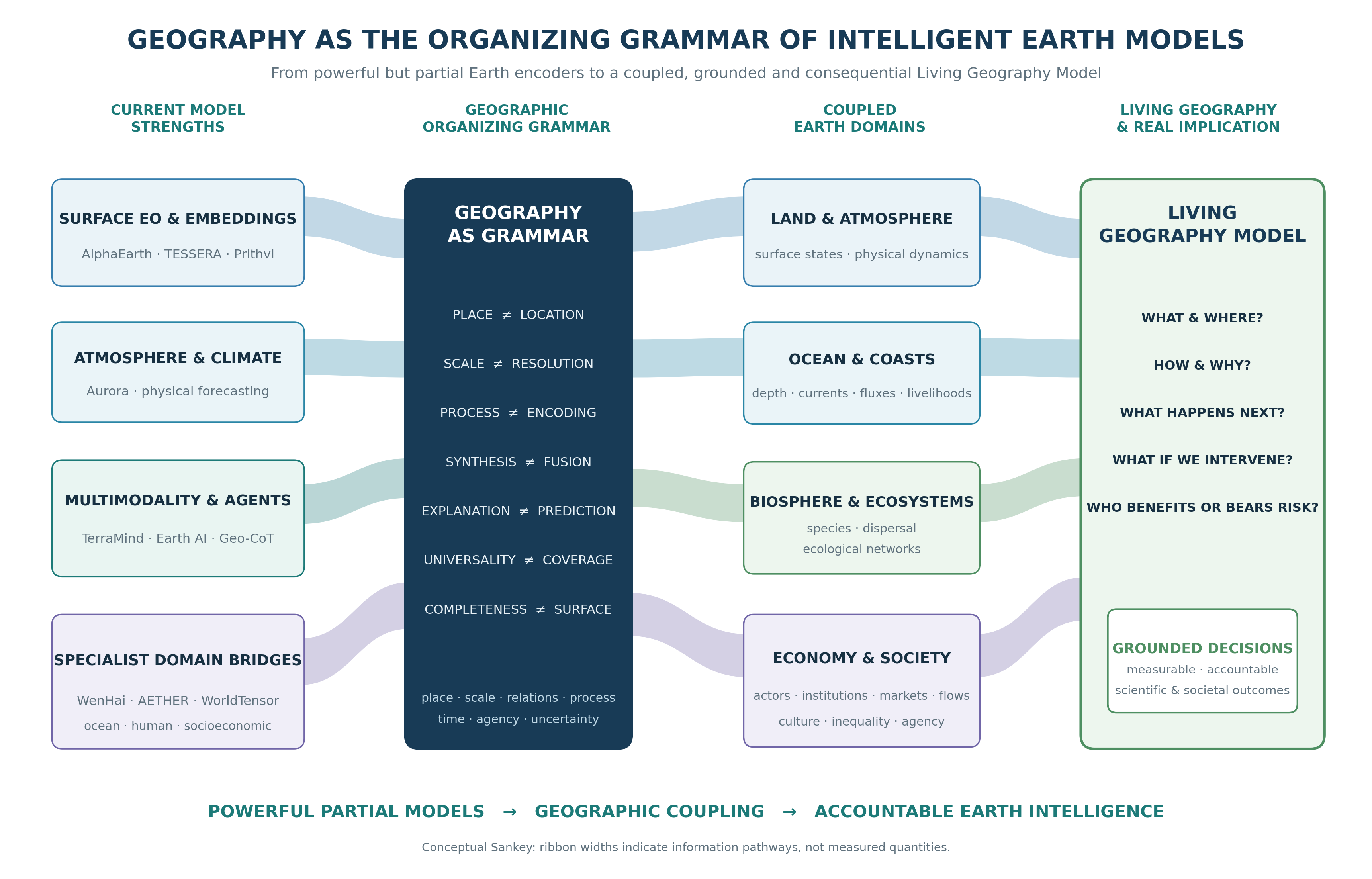}
  \caption{\textbf{Graphical abstract.} Geography provides the organizing
  grammar that connects current Earth-model strengths to coupled physical,
  biological, economic, and human domains, enabling grounded reasoning and
  accountable real-world implication.}
  \label{fig:graphical-abstract}
\end{figure*}

\section{Introduction: Geospatial Does Not Necessarily Mean Geographic}
\label{sec:introduction}

Earth observation and artificial intelligence are converging around a powerful
ambition: general computational models of the planet. Satellite archives can
now be summarized as globally consistent embeddings
\citep{brown2025alphaearth,feng2025tessera}; physical fields can be forecast by
large pretrained models \citep{bodnar2025aurora,cui2025wenhai}; observations
can be translated across modalities \citep{jakubik2025terramind,
wu2025skysense}; and language agents can coordinate geospatial tools
\citep{bell2025earthai}. These developments make fragmented observations more
accessible, reusable, and computationally tractable while reducing repeated
model development \citep{zhu2026foundations}.

The resulting systems are increasingly \emph{geospatial}, but not necessarily
\emph{geographic}. Coordinates indicate where an observation was recorded;
geography additionally asks how places are constituted, how they are connected,
which processes transform them, at what scales conclusions remain valid, and
how environmental and human systems jointly produce change
\citep{national1997rediscovering}. Geography is therefore not simply another input
modality. It provides the conceptual and computational logic for organizing
observations, relations, processes, actors, scales, evidence, and claims.

This distinction matters because the term \emph{Earth} is often spatially
extensive but ontologically selective. Foundation-model infrastructure is
strongest where observations are abundant and readily harmonized, particularly
for terrestrial surface properties, elevation, land cover, and gridded
atmospheric variables. The geographic world, however, also includes the ocean
as a three-dimensional dynamic system; organisms, communities, dispersal, and
ecological interactions; production, trade, mobility, and infrastructure
networks; and institutions, culture, inequality, and human agency
\citep{pereira2013ebv,jetz2019ebv,gao2023human,
rodriguez2026worldtensor}. A system may therefore provide global land-surface
coverage while leaving substantial parts of the geographic world structurally
unrepresented.

This does not imply that a terrestrial encoder should model culture or that a
weather model should explain migration. Specialist systems remain essential.
The problem arises when a partial observational scope is interpreted as a
complete Earth ontology, or when benchmark performance is treated as evidence
of place-based understanding and operational consequence. Spatial dependence,
aggregation sensitivity, and cross-level inference have long demonstrated that
geographic conclusions depend on spatial support, scale, and context
\citep{tobler1970law,openshaw1983maup,robinson1950ecological}. These
constraints do not disappear as representations become larger or more
multimodal.

We therefore advance a central thesis: \textbf{geography should serve as the
organizing grammar of intelligent Earth models}. We formalize this grammar
through geographic primitives places, entities, events, attributes,
relations, and flows dynamic operators processes, transitions,
trajectories, drivers, and interventions and contextual
constraints scale, spatial support, uncertainty, provenance, and validity.
Together, these elements specify how heterogeneous observations can refer to
the same evolving world without collapsing physical, biological, economic,
and social systems into a single undifferentiated representation. This
formulation extends GIScience's concern with formal geographic representation
\citep{goodchild1992giscience} and GeoAI's objective of geographic knowledge
discovery \citep{janowicz2020geoai}.

On this basis, the review identifies seven structural gaps separating
geospatial coverage from geographic intelligence and develops seven
propositions and corresponding review questions through which those gaps can
be examined. Rather than ranking specialist systems as competing universal
models, it interprets Earth-observation representations, physical forecasting
systems, human-context models, and grounded reasoning agents as complementary
components of a broader geographic architecture. The analysis culminates in
the concept of Living Geography Models: federated, continually updated,
multiscale systems that connect land, atmosphere, ocean, biosphere, economy,
and society while linking representation quality to grounded explanation,
robust transfer, accountable decisions, and demonstrable scientific and
societal implication.

\section{The Core Gap: From Geospatial Coverage to Geographic Intelligence}
\label{sec:domain-completeness}

The central gap is not a shortage of data layers or larger encoders. It is the
absence of an explicit account of what turns georeferenced pattern recognition
into geographic intelligence. Table~\ref{tab:core-gaps} states seven recurring
reductions. Each is computationally useful; none is conceptually equivalent to
the geographic capability it approximates.

\begin{table*}[t]
\centering
\caption{Seven structural gaps that define the paper's central argument.}
\label{tab:core-gaps}
\small
\begin{tabularx}{\textwidth}{@{}p{0.18\textwidth}p{0.31\textwidth}X@{}}
\toprule
\textbf{Reduction} & \textbf{Why the proxy is insufficient} &
\textbf{Geographic requirement} \\
\midrule
Location $\rightarrow$ coordinates &
Position omits history, function, meaning, and connection. &
Evolving, relational representations of place. \\
Scale $\rightarrow$ resolution &
Pixel size omits extent, support, hierarchy, and process scale. &
Cross-scale inference with aggregation and boundary sensitivity. \\
Time $\rightarrow$ encoding &
Sequence and seasonality do not identify transitions or drivers. &
Process models tested under novel forcing and regime change. \\
Multimodality $\rightarrow$ fusion &
Co-occurrence does not distinguish driver, state, intervention, and outcome. &
Typed cross-domain relations and support-aware synthesis. \\
Prediction $\rightarrow$ explanation &
Predictive fit does not establish mechanism or causal effect. &
Grounded evidence, alternatives, claim types, and uncertainty. \\
Global coverage $\rightarrow$ universality &
Worldwide extent may conceal regional, temporal, sensor, and class bias. &
Blocked transfer tests and disaggregated calibration. \\
Earth surface $\rightarrow$ geography &
Land appearance and physical fields omit much of ocean, life, economy, and
society. &
Federated land--atmosphere--ocean--biosphere--human representation. \\
\bottomrule
\end{tabularx}
\end{table*}

\subsection{The Domain-Completeness Gap}

Global surface coverage and geographic completeness are different
achievements. Current infrastructure understandably follows scalable
observations: optical and radar imagery, elevation, surface classes, and
gridded weather or climate fields. Three less visible forms of geography
remain harder to integrate.

\emph{Oceanography} requires depth-dependent temperature, salinity, currents,
mixing, bathymetry, biogeochemistry, coastal exchange, and air--sea coupling.
These states occupy a volume, not a land-surface raster. Specialist forecasting
shows that learned models can represent three-dimensional upper-ocean dynamics
\citep{cui2025wenhai}; geographic completeness requires
connecting those states to marine ecosystems, fisheries, shipping, coastal
livelihoods, hazards, and governance.

\emph{Biogeography} concerns species ranges, community composition, dispersal,
migration, succession, trophic interaction, invasion, and extinction.
Satellite representations are valuable proxies for habitat, phenology, and
disturbance \citep{feng2025tessera}, but many biological processes require field,
acoustic, genetic, and network evidence. Essential Biodiversity Variables
formalize the need for space--time--species measurements with explicit
uncertainty \citep{pereira2013ebv,jetz2019ebv}.

\emph{Economic and human geography} introduce production, labour, trade,
mobility, land tenure, institutions, culture, inequality, and agency.
Points of interest, population grids, and infrastructure layers provide useful
signals, but surface appearance cannot reliably recover lived experience,
informal systems, power, or institutional causation
\citep{gao2023human,kwan2016algorithmic}. Human geography therefore contributes
theories of place and inequality, contextual validation, participation, and
ethical limits---not merely additional features.

This gap should be addressed by a federation of specialist land, atmosphere,
ocean, ecological, economic, and human models. Their geometries and safeguards
need not be collapsed into one raster. They must instead share place
identities, spatial and temporal support, relations, provenance, and
uncertainty. Geography supplies that connective logic
\citep{zhu2026foundations}.

\subsection{The Implication Gap}

A second gap separates technical progress from real consequence. A reusable
embedding or benchmark gain becomes consequential only after it survives
geographic shift, represents a decision-relevant variable, reaches a user at
the required time, communicates uncertainty, fits an institutional workflow,
and improves a measured outcome. Evidence remains thin across this entire
chain. PANGAEA shows that foundation models do not consistently dominate
supervised baselines under common evaluation settings
\citep{marsocci2024pangaea}; a large audit finds substantial disagreement
between nominally similar model--benchmark--protocol results
\citep{corley2026sota}; and EarthShift documents degradation under geographic,
temporal, scale, and sensor shifts \citep{doerksen2026earthshift}.

The implication gap is therefore not evidence that GeoAI lacks value. It shows
that most evaluations stop at representation or prediction. Deployment is
usually an out-of-distribution and socio-technical problem: a map must align
with thresholds, calendars, costs, ownership, correction, and accountability.
Operational studies accordingly emphasize adaptation, data quality,
stakeholder integration, and empirical testing
\citep{koldasbayeva2024challenges,butsko2025worldcereal,
ghamisi2025sdg}. A geographic model should be judged not only by what it maps,
but by whether its use produces a defensible improvement for particular
places, populations, and ecosystems.

\section{Geography as an Organizing Grammar}
\label{sec:grammar}

A vocabulary of georeferenced layers does not become geography merely through
fusion. A grammar specifies what entities exist, how they relate, which
processes transform them, how scale changes meaning, which actors exert
agency, and how uncertainty constrains a claim. The metaphor is useful because
it becomes a computational design contract rather than a rhetorical analogy.

\begin{figure*}[t]
  \centering
  \includegraphics[width=0.92\textwidth]{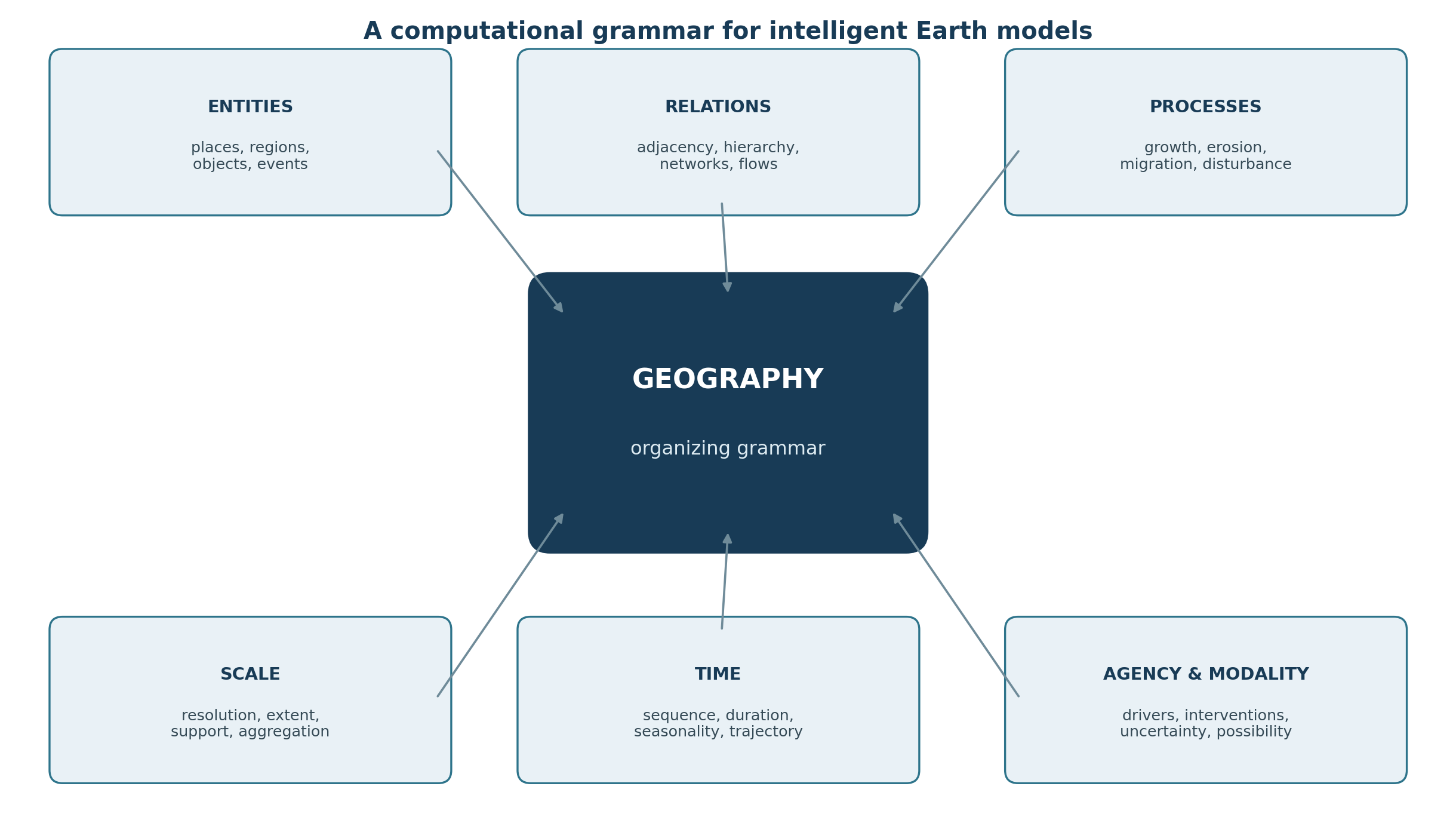}
  \caption{\textbf{Geography as a computational grammar.} Entities and places
  are organized through relations, processes, scale, time, agency, and
  uncertainty, turning observations into evidence-bearing geographic claims.}
  \label{fig:grammar}
\end{figure*}

Place is experienced, contextual, and relational
\citep{tuan1977space,agnew2011space,massey1994space}. Spatial dependence means
that samples are not exchangeable simply because they are stored independently
\citep{tobler1970law}. Scale can alter both pattern and explanation
\citep{levin1992scale}; aggregation and zoning can alter statistical
relationships \citep{openshaw1983maup}. These properties are constitutive of
geographic inference, not optional metadata.

Let the grammar be the typed system
\begin{equation}
  \Gamma=(\mathcal{V},\mathcal{R},\mathcal{P},\mathcal{S},
  \mathcal{T},\mathcal{A},\mathcal{U}),
  \label{eq:grammar}
\end{equation}
where $\mathcal{V}$ contains places, entities, and events; $\mathcal{R}$
contains adjacency, containment, connectivity, dependence, and flow;
$\mathcal{P}$ contains transition processes; $\mathcal{S}$ and $\mathcal{T}$
define spatial and temporal support; $\mathcal{A}$ represents natural and
human agency; and $\mathcal{U}$ records uncertainty and provenance. A model is
geographic to the extent that its inferences preserve constraints among these
types: a flow connects admissible entities, an event has a valid time and
support, a process operates at a defensible scale, and an explanation states
evidence, alternatives, and uncertainty.

This formulation supports modularity. An ocean model may use depth coordinates,
an atmosphere model pressure levels, an ecological model a species network,
and an economic model a trade graph. Shared identifiers and typed relations
let these systems refer to the same changing places without forcing every
domain onto one grid. Grammar is thus an interoperability layer, a reasoning
structure, and an epistemic constraint.

\section{Seven Propositions for Geographic Intelligence}
\label{sec:propositions}

The following propositions are designed as citable, contestable, and testable
claims. Together they prevent a recurring error in which a convenient
computational proxy is treated as equivalent to the geographic concept it only
partially measures.

\proposition{1: Location Is Not Place}

Location specifies position; place is constituted by environmental character,
history, function, meaning, institution, and connection. This distinction is
found across humanistic, political, and relational accounts of place
\citep{tuan1977space,agnew2011space,massey1994space}. Coordinates and location
embeddings can provide powerful priors, but they cannot alone distinguish why
apparently similar landscapes function differently. A geographic model should
represent place as an evolving relational entity rather than merely a sample
indexed by latitude and longitude.

\textbf{Testable implication.} Evaluate transfer between visually similar but
functionally different places; retrieve places from relational rather than
visual similarity alone; and require explanations to use locally valid
environmental, historical, and institutional evidence.

\proposition{2: Resolution Is Not Scale}

Spatial resolution describes measurement granularity. Geographic scale also
includes extent, support, hierarchical level, temporal duration, and the scale
at which a process operates. Pattern and explanation can change with scale
\citep{levin1992scale}. Resampling the same image to multiple pixel sizes does
not establish scale-aware reasoning, and changing aggregation or zoning can
alter inferred relationships \citep{openshaw1983maup}. Aggregate relationships
also need not describe individuals \citep{robinson1950ecological}.

\textbf{Testable implication.} Measure aggregation consistency, sensitivity to
boundary definitions, cross-level inference, and recognition that a
relationship observed at field scale may reverse or disappear at regional
scale.

\proposition{3: Temporal Encoding Is Not Process Understanding}

Timestamp embeddings, temporal attention, and image stacks help models exploit
seasonality and sequence. Process understanding requires more: representing
state transitions, separating drivers from responses, distinguishing
disturbance from recurrence, and predicting how trajectories change when
forcing changes. Earth-system science has repeatedly argued for joining
data-driven learning with process knowledge rather than treating prediction
as sufficient understanding \citep{reichstein2019process,
karpatne2017theory}.

\textbf{Testable implication.} Evaluate multiple forecast horizons, transition
regimes, unseen combinations of drivers, and counterfactual interventions.
Reconstruction of an observation at time $t$ is weaker evidence than accurate
prediction of a future state under a novel forcing.

\proposition{4: Multimodality Is Not Geographic Synthesis}

A model can fuse optical, radar, climate, topographic, textual, and
socioeconomic variables while learning only statistical co-occurrence.
Geographic synthesis requires modalities to be organized by their roles in a
process: driver, state, constraint, observation, intervention, exposure, or
outcome. The need is especially clear in coupled human--Earth systems, where
physical variables alone omit the human systems that shape environmental
change and vulnerability \citep{rodriguez2026worldtensor,ipcc2022impacts}.

\textbf{Testable implication.} Evaluate missing-modality robustness, explicit
driver interventions, temporal and spatial support compatibility, and the
stability of inferred relations across held-out places.

\proposition{5: Prediction Is Not Explanation}

High predictive accuracy establishes association under an evaluation design;
it does not establish mechanism, causal attribution, or a reason suitable for
intervention. A model that forecasts crop stress from imagery and weather may
still fail to distinguish water deficit, disease, management, or sensor
artefact. Formal causal reasoning requires assumptions and evidence beyond
predictive fit \citep{pearl2009causality}. Spatial causal methods such as
geographical convergent cross mapping illustrate that even causal inference
from Earth observations requires an explicit theory, reconstruction, and
validation procedure \citep{gao2023gccm}.

\textbf{Testable implication.} Link explanations to locatable evidence,
compare alternative hypotheses, calibrate uncertainty, and label claims as
descriptive, predictive, attributive, causal, or counterfactual. Fluent
language cannot upgrade association into explanation.

\proposition{6: Global Coverage Is Not Geographic Universality}

A globally distributed training corpus can still privilege data-rich regions,
common land-cover classes, particular sensors, and institutional definitions.
PANGAEA documents geographic and protocol limitations in GFM evaluation
\citep{marsocci2024pangaea}. EarthShift finds that evaluated models degrade
under realistic geographic, temporal, scale, and sensor shifts
\citep{doerksen2026earthshift}. Spatially structured cross-validation is
therefore necessary whenever dependence makes random splitting optimistic
\citep{roberts2017cv}.

\textbf{Testable implication.} Require region-, time-, sensor-, and
scale-held-out tests, with performance disaggregated for rare classes and
data-poor environments. A global map is spatially extensive; it is not proof
of universal validity.

\proposition{7: Earth-Surface Coverage Is Not Geographic Completeness}

A model can cover every land pixel and still represent only a selected exterior
of the Earth. Terrestrial surface embeddings and atmospheric fields are
indispensable, but geographic completeness also requires the
three-dimensional ocean, the organization and movement of life, economic
networks and flows, and the institutions, meanings, and inequalities through
which places are produced \citep{cui2025wenhai,jetz2019ebv,
gao2023human,rodriguez2026worldtensor}. Specialist systems in ocean
forecasting, biodiversity monitoring, population, and urban semantics should
be recognized as complementary advances.

\textbf{Testable implication.} Audit domain coverage, test cross-domain
consistency, and validate the learned couplings among land, atmosphere, ocean,
biosphere, economy, and society. Completeness is a property of represented
domains and valid interfaces, not merely map extent.

\section{Review Design and Analytical Scope}
\label{sec:review-design}

This manuscript is a \emph{critical integrative review}, not a performance
meta-analysis. Integrative reviews are appropriate when a field contains
heterogeneous concepts, methods, and evidence that must be synthesized into a
new conceptual framework \citep{snyder2019review}. A single numerical ranking
would be unreliable here because model comparisons vary in sensors,
pretraining corpora, spatial resolution, downstream heads, tuning budgets, and
evaluation protocols \citep{marsocci2024pangaea,corley2026sota}. The review
instead asks what geographic entities, relations, processes, domains, and
claims each model family makes possible, and which requirements remain
untested.

\subsection{Search and Selection Protocol}

We used an iterative search of arXiv, Google Scholar, publisher indexes, and
reference lists through July 2026. Search terms combined \emph{geospatial
foundation model}, \emph{Earth foundation model}, \emph{Earth observation
foundation model}, \emph{Earth-system model}, \emph{geographic reasoning},
\emph{spatial reasoning}, \emph{geospatial agent}, \emph{multimodal Earth
observation}, \emph{location embedding}, \emph{ocean foundation model},
\emph{biogeography AI}, \emph{human geography}, \emph{economic geography}, and
\emph{human--Earth system}. The temporal focus was 2020--July 2026 for model
and benchmark papers. Foundational work in geography, GIScience, ecology,
causality, and spatial validation was included without a date restriction.

Selection was purposive rather than exhaustive. We retained systems that were
highly visible, introduced a distinctive model family, made broad claims about
Earth representation or reasoning, or exposed an important evaluation or
deployment issue. Model-specific factual claims were checked against the
corresponding peer-reviewed article, official preprint record, or institutional
technical report. This design supports conceptual comparison but does not
justify claims about the frequency of every feature across the complete
literature. The limitation is made explicit in Section~\ref{sec:discussion}.

\subsection{Analytical Coding Framework}

Each system is evaluated only for capabilities demonstrated in its paper.
Coding covers primary geographic domain, observation geometry, temporal
support, entities, relations, processes, agency, uncertainty, evaluation
design, and evidence of operational implication. A capability is interpreted
as demonstrated, partially demonstrated, not demonstrated, or not assessable.
The wording \emph{not demonstrated} is deliberate: absence from a paper does
not prove that a model could never support the capability, but it does mean
that the broader claim has not yet been established by available evidence.
The coding is used to identify constructive interfaces among capability
families, not to score or rank individual models.

\section{Related Work: Complementary Building Blocks}
\label{sec:landscape}

Intelligent Earth modelling has advanced through complementary capability
families. They should not be read as competing implementations of the whole of
geography. The relevant question is what each contributes to a shared
geographic state.

\subsection{Observation and Representation}

Earth-observation foundation models transform large, mostly unlabeled archives
into reusable representations. AlphaEarth Foundations provides temporally
conditioned global embedding fields \citep{brown2025alphaearth};
Prithvi-EO-2.0 uses global multi-temporal masked pretraining
\citep{szwarcman2024prithvi}; and TESSERA learns pixel-wise representations
from irregular Sentinel-1/2 time series \citep{feng2025tessera}. TESSERA is
notable for open weights, lightweight adaptation, and accessible annual 10\,m
embeddings. TESSERA v2 studies scaling and distillation across controlled
downstream experiments, strengthening the case for compact shared terrestrial
substrates \citep{feng2026tesserav2}. These systems efficiently describe
observed and latent surface state; the geographic grammar supplies explicit
place, relation, process, and claim semantics.

TerraMind extends representation learning to any-to-any generative
multimodality across Earth-observation sources \citep{jakubik2025terramind},
while SkySense++ develops large-scale factorized multimodal pretraining
\citep{wu2025skysense}. Such models broaden the observable vocabulary.
Geographic synthesis additionally requires each modality to be identified as
an observation, state, driver, constraint, intervention, or outcome.

\subsection{Dynamics, Context, and Reasoning}

Physical foundation models contribute learned transitions. Aurora spans
weather and related environmental prediction \citep{bodnar2025aurora}; WenHai
represents depth-aware upper-ocean dynamics and air--sea forcing
\citep{cui2025wenhai}. Location and human-context models supply different
information: SatCLIP encodes coordinate-associated visual priors
\citep{klemmer2023satclip}; AETHER aligns Earth embeddings with urban points
of interest \citep{liu2025aether}; and WorldTensor harmonizes environmental,
infrastructure, hazard, and socioeconomic variables
\citep{rodriguez2026worldtensor}.

A further family coordinates evidence and operations. Earth AI composes
geospatial model families and tools \citep{bell2025earthai}; Geo-CoT grounds
analytical reasoning in perceptual evidence \citep{liu2025geocot}; and
TerraLogic decomposes geospatial questions into hierarchical workflows
\citep{yan2026terralogic}. Their value is greatest when reasoning operates
over verified entities, operations, temporal support, provenance, and
uncertainty. Existing studies also show that spatial reasoning remains
sensitive to geometry representation \citep{ji2025geospatialreasoning}.

Table~\ref{tab:model-families} summarizes these systems by contribution rather
than ranking them. Benchmark initiatives and audits complement the model
families by testing transfer, protocol sensitivity, and distribution shift
\citep{marsocci2024pangaea,corley2026sota,doerksen2026earthshift}.

\begin{table*}[t]
\centering
\caption{Representative model families and their constructive role in a
geographically organized Earth model.}
\label{tab:model-families}
\small
\begin{tabularx}{\textwidth}{@{}p{0.18\textwidth}p{0.24\textwidth}
p{0.25\textwidth}X@{}}
\toprule
\textbf{Capability family} & \textbf{Representative systems} &
\textbf{Demonstrated contribution} & \textbf{Role in the grammar} \\
\midrule
Terrestrial representation &
AlphaEarth, Prithvi, TESSERA / v2 &
Reusable surface embeddings and temporal transfer &
Attributes of evolving terrestrial places \\
Multimodal learning &
TerraMind, SkySense++ &
Cross-sensor alignment, generation, and missing-modality support &
Complementary observations with typed roles \\
Physical dynamics &
Aurora, WenHai &
Atmospheric, environmental, wave, and ocean trajectories &
Process and transition modules \\
Location and human context &
SatCLIP, AETHER, WorldTensor &
Location priors, urban semantics, and coupled variables &
Place context, networks, and human--environment links \\
Grounded reasoning &
Earth AI, Geo-CoT, TerraLogic &
Tool use, evidence grounding, and workflow planning &
Query layer over entities, operations, and provenance \\
Evaluation and deployment &
PANGAEA, EarthShift, WorldCereal studies &
Comparable evaluation, shift testing, and application protocols &
Evidence from representation to decision implication \\
\bottomrule
\end{tabularx}
\end{table*}

\section{Review Questions and Evidence-Based Answers}
\label{sec:review-questions}

\reviewquestion{1: What Do Contemporary Intelligent Earth Models Learn?}

\answer They learn complementary regularities: transferable surface
representations, geophysical trajectories, location priors, cross-modal
associations, or sequences of analytical operations
\citep{brown2025alphaearth,szwarcman2024prithvi,feng2025tessera,
bodnar2025aurora,klemmer2023satclip,bell2025earthai}. The literature supports a
landscape of powerful specialists, not yet one model that integrates place,
relation, process, scale, and human--environment agency.

\reviewquestion{2: Does Multimodality Produce Geographic Synthesis?}

\answer Not automatically. Multimodal pretraining improves alignment,
generation, and mapping \citep{jakubik2025terramind,wu2025skysense}, whereas
synthesis additionally requires compatible spatial and temporal support,
typed variable roles, stable relations, and plausible responses to
intervention. Fusion is a data operation; synthesis is a geographic claim.

\reviewquestion{3: Do Temporal Models Understand Change?}

\answer They capture useful sequences, seasonality, and predictive dynamics
\citep{feng2025tessera,bodnar2025aurora,cui2025wenhai}. Process understanding
is a stronger claim: it requires transitions under novel drivers, regime
changes, disturbances, and interventions, preferably constrained by domain
knowledge \citep{reichstein2019process,karpatne2017theory}.

\reviewquestion{4: Can Language Models Supply Geographic Reasoning?}

\answer They can plan workflows, call tools, and expose reasoning steps, but
remain sensitive to geometry representation and grounding
\citep{ji2025geospatialreasoning,liu2025geocot,yan2026terralogic}. They are
best treated as interfaces over verified geographic entities, operations,
evidence, and uncertainty rather than substitutes for geographic structure.

\reviewquestion{5: Why Is Demonstrated Real-World Implication Limited?}

\answer Benchmark performance is separated from consequence by geographic
transfer, decision timing, uncertainty, institutional ownership, and measured
outcomes. Operational evidence remains uncommon, protocol choices alter
conclusions, and foundation models do not uniformly outperform local
baselines \citep{butsko2025worldcereal,marsocci2024pangaea,
corley2026sota}. The missing object is a validated chain from representation
to decision.

\reviewquestion{6: What Makes an Earth Model Genuinely Geographic?}

\answer It represents evolving places and relations across scales;
distinguishes observations, drivers, interventions, and outcomes; predicts
transitions with calibrated uncertainty; exposes provenance; and is evaluated
under geographic shift and through decisions. These requirements align with
impact-led and human-centred GeoAI analyses
\citep{koldasbayeva2024challenges,ghamisi2025sdg,gao2023human}.

\reviewquestion{7: Which Domains Remain Structurally Underrepresented?}

\answer The imbalance favours terrestrial surfaces and gridded physical
fields. Depth-aware oceans, explicit organisms and ecological networks,
economic flows, institutions, culture, informal systems, and agency remain
distributed across specialist evidence
\citep{cui2025wenhai,pereira2013ebv,jetz2019ebv,
rodriguez2026worldtensor,kwan2016algorithmic}. The constructive response is a
federation whose components share place, scale, time, relation, provenance,
and uncertainty.

\section{Toward Living Geography Models: A Research Agenda}
\label{sec:lgm}

An \emph{Intelligent Earth Model} represents, predicts, or reasons about Earth
phenomena. A \emph{Geographic Intelligent Earth Model} organizes those
capabilities through place, scale, relation, process, flow, and
human--environment interaction. A \emph{Living Geography Model} (LGM) is the
proposed mature form: a federated, continually updateable, multiscale, and
process-aware model of coupled places.

\begin{figure*}[t]
  \centering
  \includegraphics[width=0.94\textwidth]{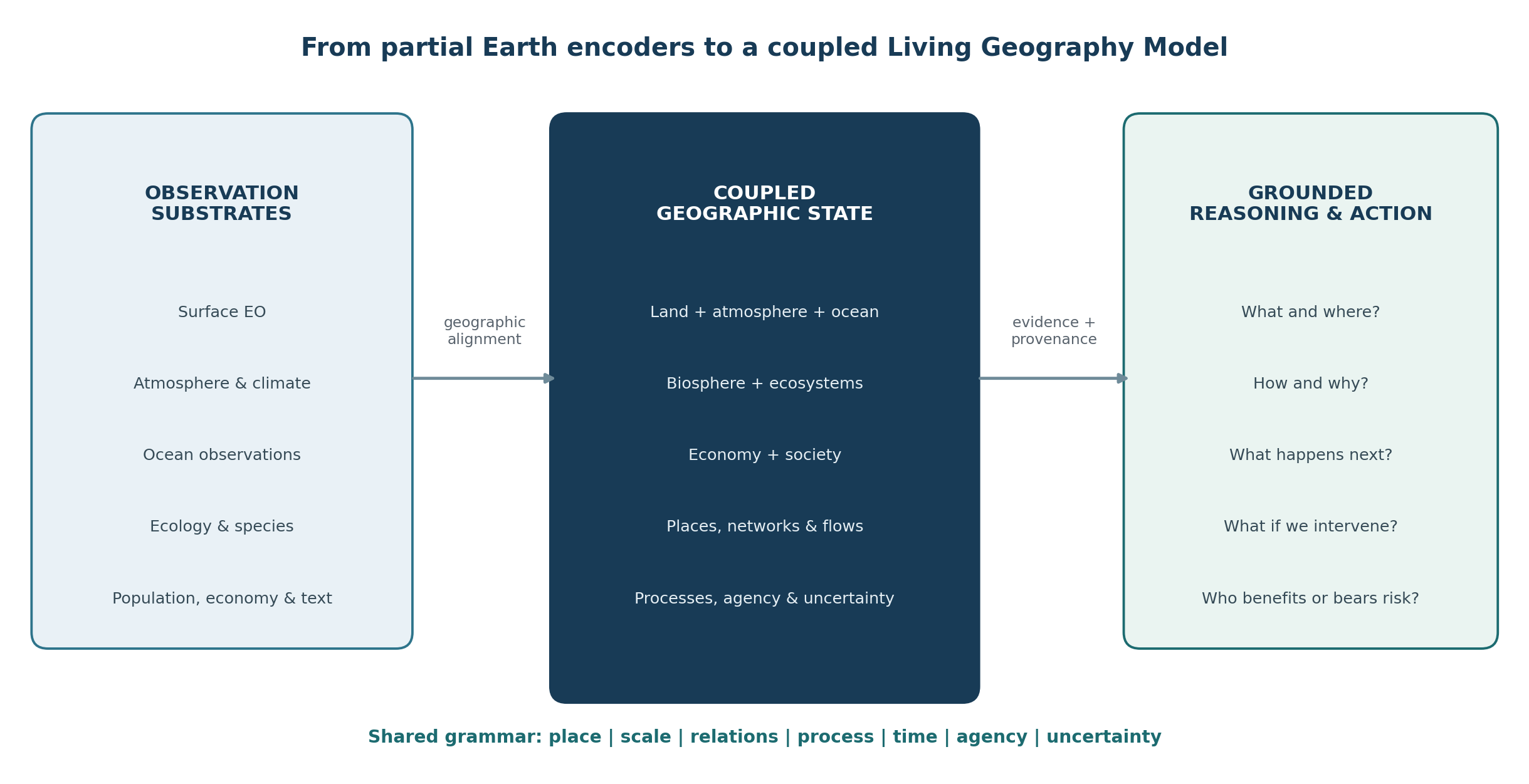}
  \caption{\textbf{Federated Living Geography architecture.} Specialist
  encoders contribute to a shared geographic state; the grammar aligns domains
  through place, scale, relations, process, provenance, and uncertainty.}
  \label{fig:lgm-architecture}
\end{figure*}

\subsection{State and Dynamics}

At time $t$ and scale $s$, define
\begin{equation}
 G_{t,s}=\{V_{t,s},E_{t,s},X_{t,s},P_{t,s},A_{t,s},U_{t,s}\},
 \label{eq:state}
\end{equation}
where $V$ denotes places, entities, and events; $E$ spatial, functional, and
flow relations; $X$ observed and latent attributes across land, atmosphere,
ocean, biosphere, economy, and society; $P$ transition processes; $A$ actors,
drivers, and interventions; and $U$ uncertainty, provenance, privacy, and
temporal validity. Domain geometries remain explicit: ocean depth, atmospheric
levels, ecological networks, and economic graphs need not be collapsed onto a
land raster.

Future states are represented as
\begin{equation}
 p(G_{t+1:t+h,s'}\mid G_{\leq t,\mathcal S},D_{t:t+h},
 \operatorname{do}(I),\Gamma),
 \label{eq:dynamics}
\end{equation}
where $\mathcal S$ contains relevant scales, $D$ external drivers, $I$ an
intervention, and $\Gamma$ the grammar in Equation~\ref{eq:grammar}. The
intervention notation is an aspiration rather than permission to infer
causality: counterfactual claims still require identification assumptions and
valid evidence \citep{pearl2009causality}.

\subsection{Architecture and Evaluation}

An LGM requires five linked capabilities: (1) multimodal observation encoders
with measurement support; (2) federated domain modules that preserve specialist
geometry and safeguards; (3) multiscale place memory containing topology,
networks, and flows; (4) cross-domain transition models constrained where
possible by science \citep{reichstein2019process,karpatne2017theory}; and
(5) a grounded query and governance layer that records evidence, claim type,
uncertainty, access, revision, and redress.

Evaluation should be organized by geographic questions---what and where, how
connected, what changed, why, what next, what if, at what scale, and for
whom---rather than only application labels. Table~\ref{tab:lgm-evaluation}
maps the paper's propositions to compact tests.

\begin{table*}[t]
\centering
\caption{Core evaluation dimensions for Living Geography Models.}
\label{tab:lgm-evaluation}
\small
\begin{tabularx}{\textwidth}{@{}p{0.18\textwidth}p{0.39\textwidth}X@{}}
\toprule
\textbf{Dimension} & \textbf{Question} & \textbf{Indicative test} \\
\midrule
Place & Are visually similar but functionally different places distinguished? &
Held-out place transfer and relational retrieval \\
Scale & Does inference survive aggregation and boundary change? &
Cross-scale consistency and MAUP sensitivity \\
Relation and process & Are topology, flows, transitions, and novel drivers
represented? & Graph, conservation, trajectory, and regime tests \\
Completeness and coupling & Are absent domains identified and coupled states
coherent? & Domain audit, lag, alignment, and interface tests \\
Explanation & Is each claim tied to evidence, alternatives, and inference type? &
Grounding, provenance, and counterfactual validity \\
Universality & Does skill persist across place, time, sensor, and scale? &
Blocked OOD tests with disaggregated calibration \\
Implication & Does model use improve a real decision or outcome? &
Prospective institutional or field evaluation \\
\bottomrule
\end{tabularx}
\end{table*}

Geographic leakage must be controlled. Neighbouring observations assigned
across training and test sets can make generalization estimates optimistic
\citep{roberts2017cv}. Benchmarks should publish exact preprocessing, tuning,
spatial and temporal splits, local baselines, and uncertainty; foundation-model
status is not itself a baseline \citep{marsocci2024pangaea,corley2026sota}.

\subsection{A Feasible Research Programme}

The agenda should proceed through interfaces rather than one planetary
training run. First, establish reproducible land, atmosphere, ocean,
biodiversity, and human-state modules with metadata and uncertainty. Second,
align pixels, fields, watersheds, habitats, settlements, administrative units,
networks, and ocean volumes through explicit relations. Third, test tractable
couplings---for example rainfall--soil--crop--management--market or
river--coast--fishery--livelihood---under distribution shift and alternative
drivers. Fourth, add grounded queries and prospective evaluation with
institutions and communities. Success is not immediate completeness; it is
evidence that a common grammar connects domains without erasing their scales,
causal structures, actors, or safeguards.

\section{Ontology, Knowledge, and Accountability}
\label{sec:philosophy}

Every Earth model contains an ontology. Raster systems foreground continuous
surface properties; administrative data foreground official boundaries; and
mobility records foreground connected populations. A geographic grammar makes
these choices inspectable: which places, actors, relations, and processes are
represented, and which remain invisible? This is consequential for informal
settlements, customary territories, smallholder fields, seasonal populations,
and processes that do not align with convenient grids
\citep{agnew2011space,massey1994space}.

The grammar also disciplines knowledge claims. Direct observation, inferred
state, prediction, attribution, causal inference, and counterfactual
simulation rest on different evidence. Causality requires explicit assumptions
\citep{pearl2009causality}; spatial dependence and spillover add geographic
complications \citep{gao2023gccm}. Each output should therefore state claim
type, support, temporal validity, provenance, and uncertainty.

Scale and representation are political as well as technical. Aggregation can
hide deprivation, minority land uses, ecological patches, and unequal
exposure \citep{openshaw1983maup,robinson1950ecological}. Accountable models
need plural place descriptions, privacy, local correction, disaggregated
errors, versioned state, monitoring, and redress
\citep{kwan2016algorithmic,gao2023human}. Where evidence is inadequate, a
system should return a bounded descriptive answer or refuse causal language.
These are properties of the model, not optional interface features.

\section{Discussion: Scope, Composition, and Falsifiability}
\label{sec:discussion}

The proposed grammar does not require every specialist model to implement
every capability. It instead limits claims to demonstrated scope. A
coordinate-aware classifier is not thereby a place model; temporal attention
is not process explanation; multimodal prediction is not whole-Earth
synthesis. This discipline complements proposals for multimodal,
scale-aware, uncertainty-aware, and physically informed Earth foundation
models \citep{zhu2026foundations}.

The same framing explains how partial models can form a coherent system.
Surface encoders, physical dynamics, ecological modules, human and economic
models, geographic graphs, and reasoning agents can remain specialized. What
makes their composition geographic is shared identity, scale, temporal
support, relation, provenance, uncertainty, and inference semantics. This
constructive middle position avoids both treating global representations as
complete Earth models and dismissing valuable systems because their intended
scope is narrower.

The thesis is falsifiable. It would be weakened if coordinate- and
time-conditioned encoders consistently passed place-difference, MAUP,
topology, process-intervention, domain-completeness, and decision tests without
explicit geographic structure. It would also be weakened if federated
interfaces introduced more error, incoherence, cost, or governance risk than
they resolved. Evidence that typed relations, process roles, and scale support
improve transfer, causal discrimination, calibrated reasoning, or outcomes
would support it.

This review is integrative rather than exhaustive. Rapid model evolution,
unequal reporting, and protocol variation constrain comparison
\citep{corley2026sota}. The proposed architecture is a research agenda, not an
empirically validated universal system. Critical, feminist, Indigenous, and
decolonial geographic traditions also deserve deeper engagement than space
permits here.

\section{Conclusion: From Earth Representation to Living Geography}
\label{sec:conclusion}

Earth models now summarize observations, transfer representations, forecast
physical fields, combine sensors, and coordinate tools at unprecedented scale.
Yet geospatial extent does not guarantee geographic intelligence; surface
coverage does not guarantee domain completeness; and benchmark progress does
not guarantee real implication.

Geography supplies an organizing grammar. Places and events are entities;
relations are syntax; processes are verbs; time supplies tense; scale supplies
context; natural and human forces establish agency; and uncertainty defines
modality. The resulting Living Geography Model is not one network that
contains all knowledge. It is a federated, updateable system whose components
refer to the same evolving world and whose claims remain grounded, scoped, and
testable. This offers a path from powerful partial models toward accountable
Earth intelligence that connects land, atmosphere, ocean, life, economy, and
society---and demonstrates consequence in the places where it is used.

\section*{Data and Code Availability}
This article is a conceptual critical review and reports no new empirical
dataset or trained model. The arXiv source package contains the complete
\LaTeX{} manuscript, bibliography, and original conceptual figures.

\section*{Acknowledgements}
The authors thank the Earth-observation, Earth-system science, geography, and
GeoAI communities whose open papers, models, data products, and benchmarks made
this synthesis possible.

\clearpage
\balance
\bibliographystyle{unsrtnat}
\bibliography{refs}

\end{document}